\documentclass[twocolumn,superscriptaddress,prl]{revtex4}    
\usepackage{graphicx,amsmath,mathrsfs}
\usepackage{amssymb}
\usepackage{amsthm,multirow}
\usepackage{bm}
\usepackage{subfigure,xcolor}   

\usepackage{color}

\def\bc{\begin{center}}
\def\ec{\end{center}}

\newcommand{\bs}[1]{\boldsymbol{#1}}

\newcommand{\up}{\uparrow}
\newcommand{\dw}{\downarrow}
\newcommand{\pd}{{\phantom{\dagger}}}

\def\ie{\emph{i.e.},\ }
\def\eg{\emph{e.g.}\ }

\begin{document}
\title{Quantum disordered insulating phase in the frustrated cubic-lattice Hubbard model}
\author{Manuel Laubach}
\affiliation{Institut f\"ur Theoretische Physik, Technische Universit\"at Dresden, 01062 Dresden, Germany}
\affiliation{Institut f\"ur Theoretische Physik, Universit\"at W\"urzburg, 97074 W\"urzburg, Germany}
\author{Darshan\,G.\,Joshi}
\affiliation{Institut f\"ur Theoretische Physik, Technische Universit\"at Dresden, 01062 Dresden, Germany}
\author{Johannes\,Reuther}
\affiliation{Dahlem\,Center\,for\,Complex\,Quantum\,Systems\,and\,Fachbereich\,Physik,\,Freie\,Universit\"at\,Berlin,\,14195 Berlin,\,Germany}
\affiliation{Helmholtz-Zentrum Berlin f\"ur Materialien und Energie, 14109 Berlin, Germany}
\author{Ronny\,Thomale}
\affiliation{Institut f\"ur Theoretische Physik, Universit\"at W\"urzburg, 97074 W\"urzburg, Germany}
\author{Matthias\,Vojta}
\affiliation{Institut f\"ur Theoretische Physik, Technische Universit\"at Dresden, 01062 Dresden, Germany}
\author{Stephan\,Rachel}
\affiliation{Institut f\"ur Theoretische Physik, Technische Universit\"at Dresden, 01062 Dresden, Germany}


\begin{abstract}
In the quest for quantum spin liquids in three spatial dimensions (3D), we study the half-filled Hubbard model on the simple cubic lattice with hopping processes up to third neighbors. Employing the variational cluster approach (VCA), we determine the zero-temperature phase diagram: In addition to a paramagnetic metal at small interaction strength $U$ and various antiferromagnetic insulators at large $U$, we find an intermediate-$U$ antiferromagnetic metal. Most interestingly, we also identify a non-magnetic insulating region, extending from intermediate to strong $U$.
Using VCA results in the large-$U$ limit, we establish the phase diagram of the corresponding $J_1$--$J_2$--$J_3$ Heisenberg model. This is qualitatively confirmed -- including the non-magnetic region --  using spin-wave theory.
Further analysis reveals a striking similarity to the behavior of the $J_1$--$J_2$ square-lattice Heisenberg model, suggesting that the non-magnetic region may host a 3D spin-liquid phase.
%
\end{abstract}


\maketitle


{\it Introduction}.---
Quantum spin liquids (QSLs), characterized by the absence of long-range order down to the lowest temperatures, are amongst the most fascinating states of matter \cite{anderson73mrb153,balents10n199}.
Their elementary excitations often 
display {\it fractionalization}, resulting in unconventional quasiparticles with exotic properties.
QSL states typically emerge from a combination of frustration, \ie the competition of different exchange couplings, and quantum fluctuations, both acting to suppress symmetry-breaking order.
Given that fluctuation effects in lattice systems weaken with increasing coordination number, the majority of works in the field have focussed on systems in two dimensions (2D), while 3D spin liquids are scarce because they are harder to stabilize. A notable exception are spin-ice pyrochlores which, however, are {\em classical} spin liquids\,\cite{bramwell-01s1495}.
Concerning 3D QSLs, some of the few promising experimental candidates are the hy\-per\-ka\-go\-me compound Na$_4$Ir$_3$O$_8$\,\cite{okamoto-07prl137207} and the pyrochlore system Pr$_2$Ir$_2$O$_7$\,\cite{machida-10n210}, but also Tb$_2$Ti$_2$O$_7$\,\cite{gardner-99prl1012}, Yb$_2$Ti$_2$O$_7$\,\cite{applegate-12prl097205}, and FeSc$_2$S$_4$\,\cite{krimmel-05prl237402} might fall into this category.
%

Theoretical work is mainly limited to a few exactly soluble spin models\,\cite{hermele-04prb064404,castelnovo-08prb155120,normand-14prl207202,hermanns-15prl157202}, quantum dimer models\,\cite{moessner-03prb184512,raman-05prb064413}, and spin-liquid states predicted using parton constructions\,\cite{pesin-10np376,savary-12prl037202,wang-15arXiv:1505.03520,senthil-03prl216403}.
In contrast, results from numerical simulations, proven to be extremely powerful in the 2D case, are largely missing for 3D. The main reason is computational complexity, which severely limits the application of exact-diagonalization and density-matrix renormalization-group methods. While quantum Monte Carlo techniques are available in principle\,\cite{hirsch87prb1851}, models of frustrated electrons typically suffer from the sign problem.
Relevant models in 3D have therefore been studied using approximate methods such as
spin-rotation-invariant Green's functions\,\cite{richter-15arXiv}, dynamical mean-field theory (DMFT)\,\cite{fuchs-11prl030401,fuchs-11prb235113} and its cluster extensions\,\cite{maier-05rmp1027,go-12prl066401} as well as cluster perturbation theory\,\cite{witczakKrempa-14prl136402}.
However, obtaining reliable results concerning magnetic ground states has proven notoriously difficult.

In this Rapid Communication, we partially fill this gap by generalizing the VCA\,\cite{potthoff03epjb335} to 3D and applying it to the Hubbard model with longer-ranged hoppings on the cubic lattice. We are able to compute zero-temperature phase diagrams and single-particle spectra for arbitrary degree of frustration and for arbitrary interaction strength.
We determine the location of the metal-to-insulator transition (MIT) as well as the boundaries of various types of collinear antiferromagnetic order. Most remarkably, we identify an insulating region extending from intermediate to strong interactions which is devoid of magnetic order, rendering it a spin-liquid candidate. We investigate the large-$U$ limit using VCA and compare this to spin-wave theory for the corresponding frustrated Heisenberg model, with good qualitative agreement. Based on 
striking parallels to the behavior of the frustrated square-lattice Heisenberg model we suggest that the cubic-lattice non-magnetic insulating region may host both a 3D QSL and a valence-bond solid.

\begin{figure*}
\includegraphics[scale=0.73]{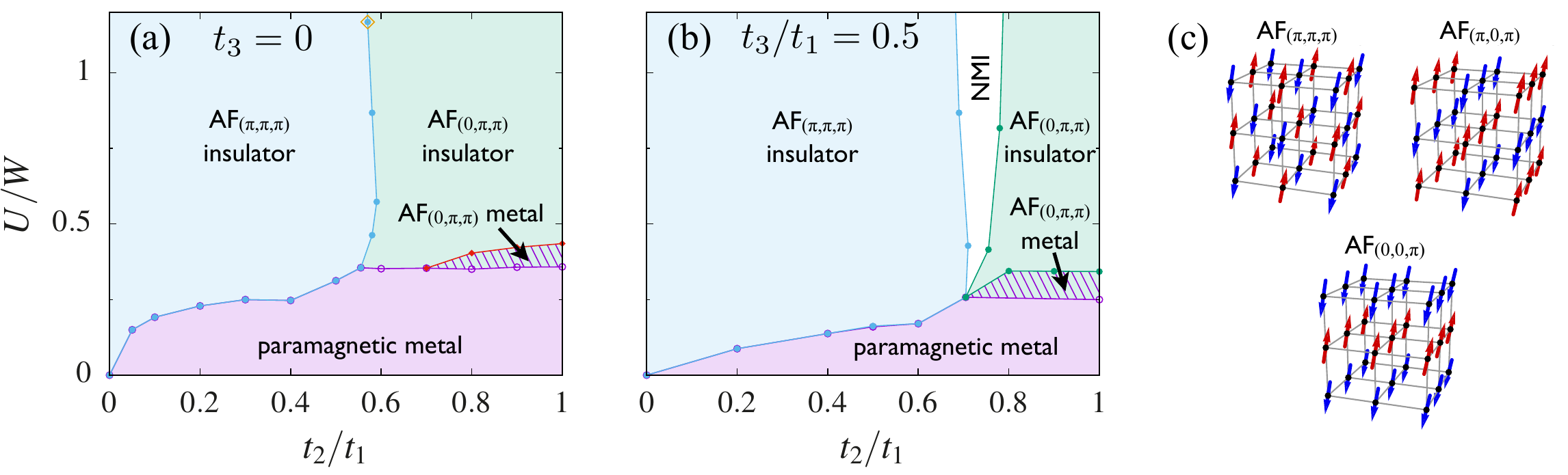}
%
\caption{
Phase diagrams of the cubic-lattice Hubbard model, plotted as function of $U/W$ and $t_2/t_1$ at fixed $t_3=0$ (a) and $t_3/t_1=0.5$ (b), with $W$ being the bandwidth of the dispersion at $U=0$. The small-$U$ paramagnetic metal is destroyed in favor of antiferromagnetic insulating phases, for labels see text. In (b) we also find a Mott-insulating regime without magnetic order (white region) which persists up to large $U$. Finally, the shaded region in (a) and (b) is an antiferromagnetic metal.
(c) Illustration 
of the three magnetic orders.
%
}
\label{fig:Hubbard}
\end{figure*}

{\it Model}.---We study the Hubbard model with hoppings up to third neighbors on the simple cubic lattice, with
\begin{equation}\label{ham-hubb}
\begin{split}
\mathcal{H} =& -t_1 \sum_{\langle ij \rangle} c_{i\sigma}^\dag c_{j\sigma}^\pd -
t_2 \sum_{\langle\!\langle ij \rangle\!\rangle} c_{i\sigma}^\dag c_{j\sigma}^\pd -
t_3 \sum_{\langle\!\langle\!\langle ij \rangle\!\rangle\!\rangle} c_{i\sigma}^\dag c_{j\sigma}^\pd\\
& \quad\quad + U \sum_i n_{i\up}n_{i\dw} 
\end{split}
\end{equation}
in standard notation. The chemical potential is chosen such that half-filling is ensured. Throughout the paper, we assume $t_1=1$ and $0\leq t_{2,3}\leq 1$. For $t_2=0$ the spectrum is particle-hole symmetric.
Non-zero $t_2$ introduces frustration in the sense of destructive interference of hopping processes; this is also apparent in the large-$U$ limit where $t_2$ generates a frustrating exchange coupling.


{\it Variational Cluster Approximation}.---VCA is a quantum cluster approach, designed to compute single-particle properties of interacting many-body systems\,\cite{potthoff03epjb335}. One first solves a small cluster exactly and derives the corresponding full Green's function. Using the framework of self-energy functional theory\,\cite{potthoff03epjb429}, one obtains the full Green's function of an infinite lattice which is covered by the clusters, and the individual clusters are coupled by hopping terms only, but in a self-consistent, variational scheme. The latter step represents a significant approximation to the full many-body problem, while the method still includes spatial quantum correlations. Embedded into a grand-canonical ensemble, the stationary configuration with lowest potential $\Omega$ is found by varying the chemical potential as well as all single-particle parameters of the reference cluster. Magnetic instabilities can be investigated by means of Weiss fields. VCA was first introduced for 1D systems\,\cite{potthoff-03prl206402} and later generalized to 2D\,\cite{potthoff03epjb335}.
In recent work, we have demonstrated that VCA is capable of detecting\,\cite{rachel-15prl167201} or rejecting\,\cite{PhysRevB.90.165136} non-magnetic insulator phases which might correspond to spin liquids.
For further VCA details we refer to Refs.\,\onlinecite{potthoff03epjb335} and \onlinecite{PhysRevB.90.165136}.

\begin{figure}[b]
\centering
\includegraphics[scale=0.58]{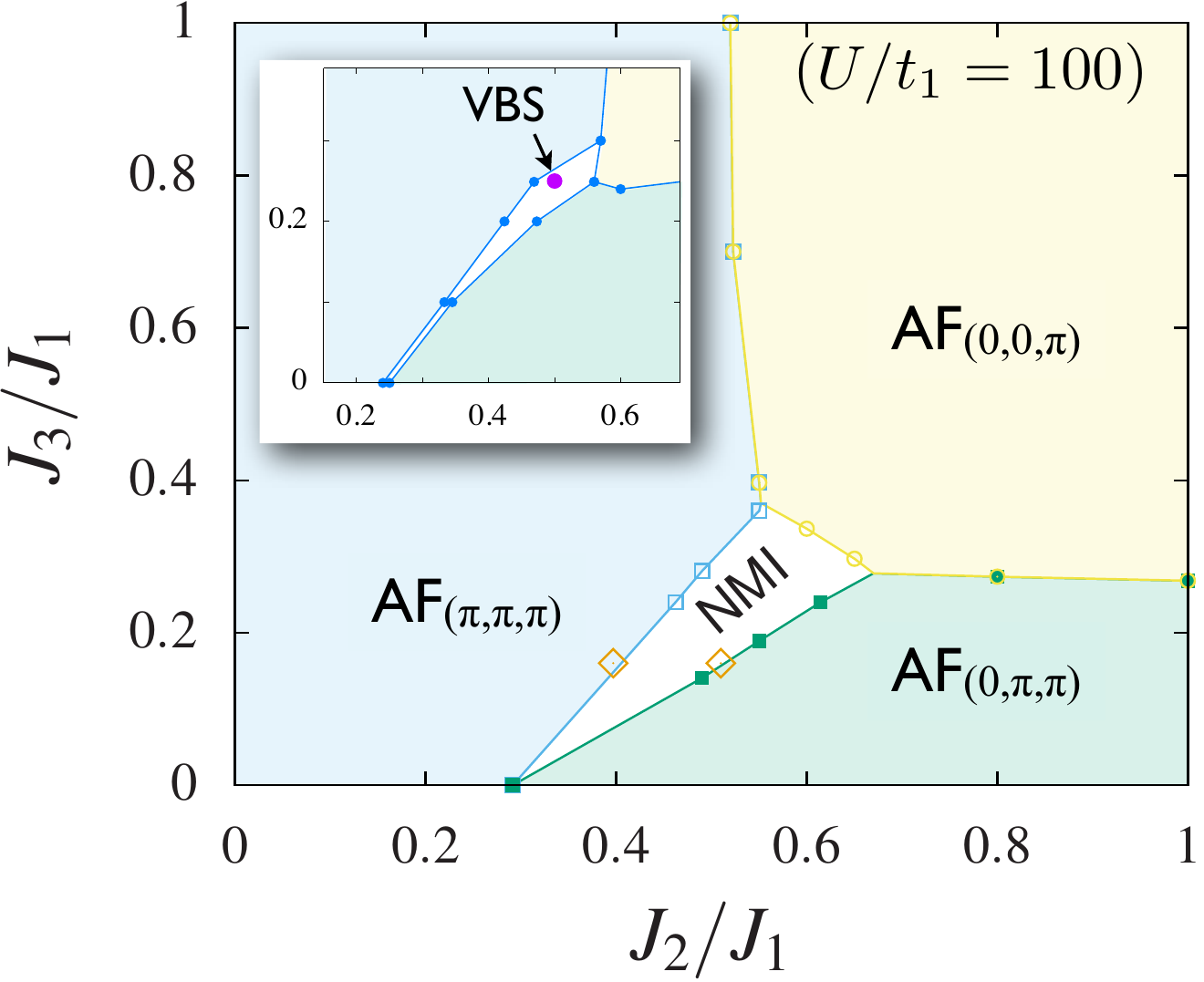}
\caption{Phase diagram of the $J_1$--$J_2$--$J_3$ cubic-lattice Heisenberg model, obtained by VCA calculations for the Hubbard model at $U/t_1=100$ where $J_{2,3}/J_1=t_{2,3}^2/t_1^2$,  for labels see text. Inset: Phase diagram as obtained within linear spin-wave theory. The purple dot within the NMI region indicates a stable solution of a columnar VBS, see text.}
\label{fig:j1j2j3-phasedia}
\end{figure}

Here we generalize and apply VCA to 3D for the first time. We use clusters consisting of $2^3=8$ lattice sites (cube) or $3\times 2^2=12$ lattice sites (rectangular cuboid) to compute the full Green's function of the Hubbard model \eqref{ham-hubb}. In fact, most data were obtained using the cubic 8-site cluster, and a few selected data points were double-checked with the 12-site cluster: the 12-site-cluster results are shown as orange diamonds in Fig.\,\ref{fig:Hubbard}\,(a) and in Fig.\,\ref{fig:j1j2j3-phasedia}; deviations turned out to be negligible.
As shown earlier\,\cite{PhysRevB.90.165136}, it is crucial to vary {\it all} the single-particle parameters, \ie not only the chemical potential but also $t_{1,2,3}$, and we follow this protocol here.

\begin{figure*}
\includegraphics[scale=0.64]{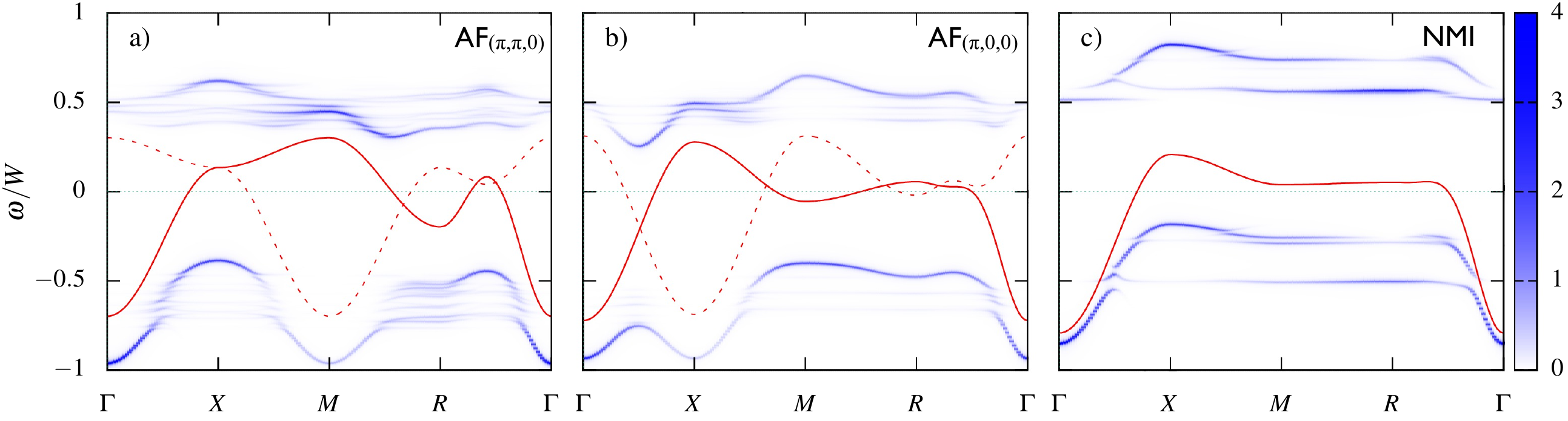}
\caption{
Intensity plot of the single-particle spectral function $A(\bs{k},\omega)$ along the momentum-space path $\Gamma$--$X$--$M$--$R$--$\Gamma$ with $\Gamma=(0,0,0)$, $X=(\pi,0,0)$, $M=(\pi,\pi,0)$, and $R=(\pi,\pi,\pi)$. (a) AF$_{(\pi,\pi,0)}$ at parameters $t_1=t_2=1$, $t_3=0$, and $U=W=24$. (b) AF$_{(\pi,0,0)}$ for $t_1=t_2=t_3=1$ and $U=W=36$. (c) NMI for $t_2/t_1=0.73$, $t_3/t_1=0.5$, and $U=W=23.68$. For comparison, the non-interacting dispersion is shown (solid, red); in (a) and (b), also the $\vec Q$-shifted (backfolded) non-interacting dispersion is shown (dashed, red).
}
\label{fig:akw}
\end{figure*}

{\it Results}.---Using VCA we compute the Green's function for each parameter triple ($t_2/t_1$, $t_3/t_1$, $U/t_1$) from which we extract the single-particle spectral function $A(\bs{k},\omega)$. Its momentum integral readily allows to distinguish between metallic and insulating phases. In addition, we used Weiss fields to test the presence of the antiferromagnetic (AF) orders known from the classical $J_1$--$J_2$--$J_3$ Heisenberg model: AF$_{(\pi,\pi,\pi)}$, AF$_{(0,\pi,\pi)}$, and AF$_{(0,0,\pi)}$ where the subscript denotes the ordering vector $\vec Q$ [see Fig.\,\ref{fig:Hubbard}\,(c)]. These orders are also referred to as G-type, C-type, and A-type AF, respectively\,\cite{wollan-55pr545}. Note that the latter two possess a discrete broken $\mathbb{Z}_3$ symmetry in addition to the broken SU(2) symmetry.
Notably, all these magnetic orders are collinear and display a two-site magnetic unit cell.

Two representative zero-temperature phase diagrams are shown in Figs.\,\ref{fig:Hubbard}\,(a,b).
At small $U$ we find the expected paramagnetic metal, descending from the non-interacting limit, with the exception of the case $t_{2}=0$ where the AF$_{(\pi,\pi,\pi)}$ insulator persists down to infinitesimal values of $U$ due to perfect nesting. While $t_2\neq0$ destroys the perfect nesting and thus stabilizes the small-$U$ metal, finite $t_3$ counteracts the effect of $t_2$ which is reflected in an effective reduction of the metallic phase [see Fig.\,\ref{fig:Hubbard}\,(b) for $t_3/t_1=0.5$].
At large $U$ we find AF insulating phases for most parameter values, with AF$_{(\pi,\pi,\pi)}$ (AF$_{(0,\pi,\pi)}$) being realized at small (large) $t_2/t_1$, respectively. If both $t_2$ and $t_3$ are large there is also an AF$_{(0,0,\pi)}$ insulator. While the MIT coincides with the onset of magnetism at small and moderate $t_2$ (and is thus presumably of first order), for large $t_2$ we observe instead that the MIT is preceded by a magnetic transition: The resulting intermediate AF$_{(0,\pi,\pi)}$ metal displays a Fermi surface with electron and hole pockets which shrink upon further increasing $U$ and disappear at the MIT.

The most remarkable finding is a narrow insulating region between AF$_{(\pi,\pi,\pi)}$ and AF$_{(0,\pi,\pi)}$ phases where magnetic order is absent. We found this non-magnetic insulating (NMI) region for values of the third-neighbor hopping $0<t_3/t_1 \lesssim 0.6$. It emerges near the point where the paramagnetic metal and two magnetic insulators meet and persists up to the strong-coupling limit where it can be associated with a quantum-disordered regime of local moments. Its nature will be discussed below.

Sample results for the single-particle spectrum $A(\bs{k},\omega)$, measurable in angle-resolved photoemission spectroscopy (ARPES), are shown in Fig.\,\ref{fig:akw}: AF$_{(\pi,\pi,0)}$ in panel a), AF$_{(\pi,0,0)}$ in panel b), and NMI in panel c). The interaction strength $U$ is  chosen such that $U/W=1$ in all cases. In addition to $A(\bs{k},\omega)$ we also show the non-interacting dispersion for comparison.
While the data in Figs.\,\ref{fig:akw}(a,b) correspond to states deep in the AF phase, the resulting spectrum is more complicated than a simple two-band signal that would be obtained from the unit-cell doubling of the band dispersion, indicating Hubbard-type interaction effects. We note that our $A(\bs{k},\omega)$ data do not display true lifetime broadening since we perform VCA calculation without bath sites. Nonetheless, one can observe some precursor of broad bands, \ie dense sequences of levels, which we identify as lower and upper Hubbard bands.
A remarkable feature in Fig.\,\ref{fig:akw}(c), corresponding to the NMI region, is the weak momentum dependence indicating the high degree of frustration -- this is already visible in the non-interacting dispersion.

\begin{figure}[t]
\centering
\includegraphics[scale=0.59]{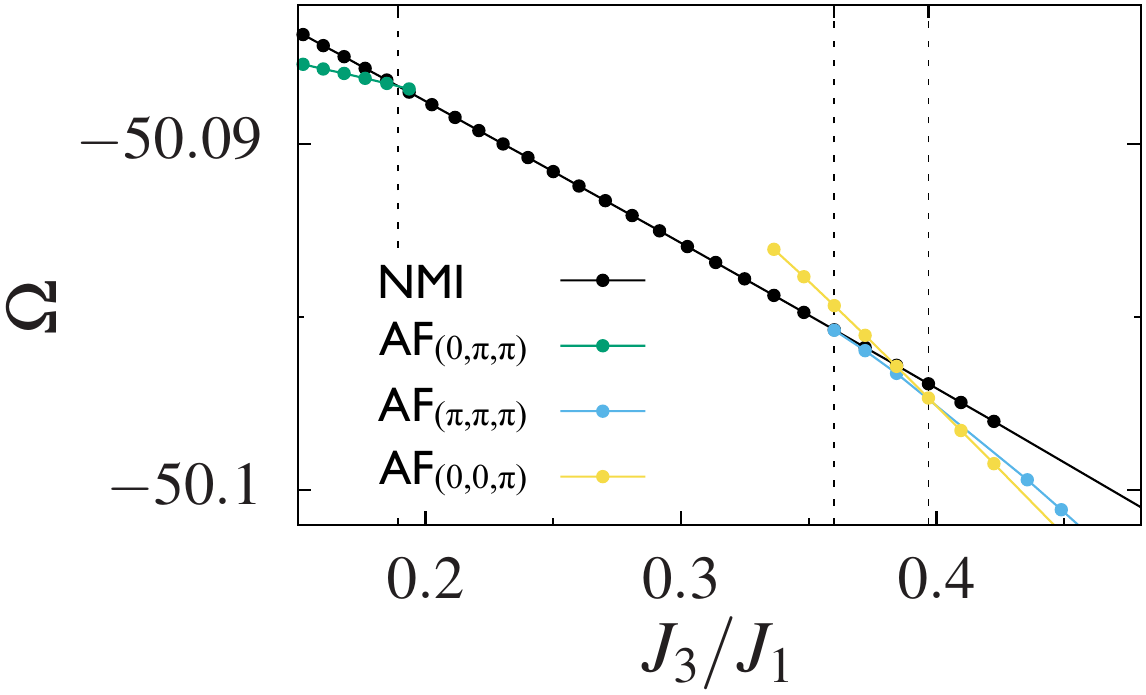} 
\caption{Grand potential $\Omega$ as a function of $J_3/J_1$ at fixed $J_2/J_1=0.55$ for $U/t_1=100$. Shown are the energies for AF$_{(\pi,\pi,\pi)}$, AF$_{(0,\pi,\pi)}$, AF$_{(0,0,\pi)}$ phase and the non-magnetic insulator regime. Ending of lines indicates the disappearance of stationary points.}
\label{fig:f_vs_t3}
\end{figure}

{\it Strong-coupling limit}.---In the limit of large on-site repulsion $U$ we expect charge fluctuations to become irrelevant. The remaining spin degrees of freedom of the Mott insulator are conveniently described using an effective Heisenberg Hamiltonian which can be obtained perturbatively in $t/U$. The present Hubbard model features real spin-independent hopping, such that every hopping process yields an AF Heisenberg coupling $J_i=4t_i^2/U$. Thus the Hamiltonian \eqref{ham-hubb} simply reduces to the $J_1$--$J_2$--$J_3$ spin-1/2 Heisenberg model. Consequently, by fixing $U$ to large enough values (here $U/t_1=100$), our Hubbard-model calculation within VCA will yield the $J_1$--$J_2$--$J_3$ quantum phase diagram which has been derived here for the first time [see Fig.\,\ref{fig:j1j2j3-phasedia}]. Since the corresponding phase diagram at $U/t_1=30$ shows only minimal deviations compared to Fig.\,\ref{fig:j1j2j3-phasedia}, we conclude that we are deep in the Mott phase and the Heisenberg description is justified. In Fig.\,\ref{fig:f_vs_t3} we have exemplarily shown the grand potential as a function of $J_3/J_1$ at fixed $J_2/J_1=0.55$ for different magnetic solutions (AF$_{(\pi,\pi,\pi)}$, AF$_{(0,\pi,\pi)}$, and AF$_{(0,0,\pi)}$) and the paramagnetic solution (NMI). The stable solution with lowest value 
$\Omega$ corresponds to the ground state as shown in the phase diagram Fig.\,\ref{fig:j1j2j3-phasedia}.

While there is no alternative method available to test our 3D VCA results at intermediate $U$ in the frustrated regime, the large-$U$ limit lends itself to comparisons with approximate solutions of the Heisenberg model. The stability of the AF phases is most naturally tested using linear spin-wave theory (LSWT). This has been applied to the $J_1$--$J_2$ model on the cubic-lattice in Ref.\,\onlinecite{majumdar-10jsp714}, and we extend this to include $J_3$ here. LSWT qualitatively confirms the VCA phase diagram Fig.\,\ref{fig:j1j2j3-phasedia}, where the LSWT result is shown as inset. Most importantly, it confirms the existence of the non-magnetic region as found within VCA.
While the predictive power of LSWT may be questioned in principle\,\cite{footnote-NLSWT}, we point out remarkable similarities to the well-studied case of the $J_1$--$J_2$ model on the square lattice\,\cite{footnotej1j2j3}. Here, LSWT correctly predicts\,\cite{chandra-88prb9335} the non-magnetic region around $J_2/J_1=0.5$ which has later been confirmed numerically (see \eg Refs.\,\onlinecite{gong-14prl027201,jiang-12prb024424,hu-13prb060402,yu-12prb094407,isaev-09prb024409} and references therein). Moreover, a quantitative comparison of the phase boundaries obtained by LSWT\,\cite{chandra-88prb9335} and the recent numerical analysis\,\cite{gong-14prl027201,jiang-12prb024424,hu-13prb060402,yu-12prb094407,isaev-09prb024409} shows that LSWT predicts a non-magnetic region which is too narrow and also shifted towards the AF phase with $\vec Q=(\pi,\pi)$; remarkably, that we find the same trend on the cubic lattice when comparing LSWT and VCA data. Together, this indicates that LSWT is appropriate in benchmarking the large-$U$ VCA results\,\cite{footnote-NLSWT}.

The shape of the phase boundaries in Fig.~\ref{fig:j1j2j3-phasedia} can be rationalized as follows: Upon increasing $J_3/J_1$ the balance between the AF$_{(\pi,\pi,\pi)}$ and AF$_{(0,\pi,\pi)}$ phases is shifted to larger $J_2/J_1$ -- this is because $J_3>0$ does not compete with $J_1>0$ and thus stabilizes the AF$_{(\pi,\pi,\pi)}$ phase relative to AF$_{(0,\pi,\pi)}$. The presence of NMI is related to quantum fluctuations: While the competition of $J_1$ and $J_2$ is apparently insufficient to destroy long-range order (unlike in the square lattice), the {\em additional} fluctuation processes arising from $J_3$ succeed in suppressing order.

We finally discuss the nature of the NMI region which cannot be accessed by the present VCA calculations. 
Prominent candidate states are valence-bond solids (VBS) with broken translational symmetry and various types of spin liquids\,\cite{footnoteFiniteSize}. In contrast, long-wave\-length non-collinear spirals (not captured by VCA) appear unlikely because spirals are absent in the classical phase diagram, and quantum fluctuations typically prefer collinear instead of non-collinear states. Here we investigate the stability of a columnar VBS, with
%
%
%
%
%
a two-site unit cell and ordering vector $(\pi,0,0)$, using bond-operator theory\,\cite{sachdev-90prb9323}. In the harmonic approximation\,\cite{footnoteBondOperator}
we find that the columnar VBS is marginally stable, \ie it exists only at $(J_2/J_1, J_3/J_1)=(0.5, 0.25)$ where it is gapless -- this is the point where the three magnetic phases meet in the classical limit.
%
Again, this reveals a remarkable similarity to the $J_1$--$J_2$ square-lattice model\,\cite{chubukov-91prb12050,kotov-99prb14613,kotov-00pmb1483} where -- within harmonic approximation -- a columnar VBS state is only stable at the classical transition point $J_2/J_1=0.5$\,\cite{kotov-00pmb1483}. It is further known that inclusion of interactions stabilizes a gapped VBS state in a finite window of parameters\,\cite{chubukov-91prb12050,kotov-99prb14613}, and we expect this to apply to the cubic-lattice case as well.
While numerical studies in the past suggested either a VBS phase or a spin liquid to be realized\,\cite{jiang-12prb024424,hu-13prb060402,yu-12prb094407,isaev-09prb024409}, very recent numerical studies\,\cite{gong-14prl027201} of the $J_1$--$J_2$ square-lattice model indicate that its non-magnetic region is split into a VBS phase (at larger $J_2$ in vicinity to the collinear state) and a spin-liquid phase (at smaller $J_2$ in vicinity to the Neel state).
Based on the striking similarities between the square-lattice results and our findings for the cubic lattice we speculate that at least a part of the NMI region in Fig.\,\ref{fig:j1j2j3-phasedia} hosts a 3D QSL. %
Note that these similarities between 2D and 3D are not accidental but rather generic: at the classical transition points $(J_2/J_1, J_3/J_1)=(0.5, 0.25)$ (3D) and $J_2/J_1=0.5$ (2D) both Hamiltonians can be expressed as the sum over  $\bs{S}_{\rm tot}^2$ operators of elementary cubes (3D) or squares (2D), respectively. 


Eventually we  comment on the relevance of the putative spin liquid phase for real materials. To stabilize the non-magnetic insulating phase rather large ratios of $J_2/J_1$ and $J_3/J_1$ are required. This is, however, not an obstacle. The search for the spin-liquid phase in the analog square lattice case stimulated efforts in crystal growth. As a result, the perovskite PbVO$_3$ features $J_2/J_1\approx 0.2 \ldots 0.4$\,\cite{tsirlin-08prb092402,oka-08ic7355}. The compound BaCdVO(PO$_4$)$_2$ even reaches $J_2/J_1\approx -0.9$ (here $J_1$ is ferromagnetic)\,\cite{nath-08prb064422}; and in Li$_2$VO(Si,Ge)O$_4$ couplings as large as $J_2/J_1 \sim 12$ have been found\,\cite{rosner-02prl186405,melzi-00prl1318}. In the light of these facts the con\-sidered magnetic couplings for the cubic lattice might be identified in future experiments.


{\it Conclusion and Outlook}.---We have applied VCA to the half-filled frustrated Hubbard model on the simple cubic lattice and computed zero-temperature phase diagrams. In addition to the weak-coupling paramagnetic metals and antiferromagnetic insulators at strong coupling, we have found an antiferromagnetic metal at intermediate coupling and -- most importantly -- an extended non-magnetic, \ie quantum-disordered, insulating region. Using the VCA data at $U/t_1=100$ we have established the phase diagram of the $J_1$--$J_2$--$J_3$ spin-1/2 Heisenberg model, whose structure could be confirmed using  linear spin-wave theory. The unexpected but striking similarity to the $J_1$--$J_2$ square-lattice model suggests that the non-magnetic region hosts both a VBS\,\cite{vbs} and a 3D QSL.

Our results raise several issues which will be subject of future work: (i) Opposite-sign hopping processes, e.g., $t_2/t_1<0$, as realized in cuprate superconductors, may modify the physics at intermediate $U$. (ii) Without perfect nesting, superconductivity may occur at half filling at intermediate $U$. For the cubic lattice this can be tested using VCA. (iii) The physics of the Hubbard model \eqref{ham-hubb} away from half filling is unexplored, but may be tackled using VCA as well. (iv) Given that this work establishes VCA as a suitable method for 3D Hubbard models, its application to other 3D lattices such as hexagonal, pyrochlore, and hyperkagome promises to unravel further novel physics.


{\it Acknowledgements}.---
We acknowledge financial support
by the DFG through SFB 1143 (ML, DGJ, MV, SR) and SFB 1170 (RT),
by the ERC through the starting grant TOPOLECTRICS (ERC-StG-Thomale-336012, RT), and
by the Helmholtz association through the virtual institute VI-521 (MV, SR).
We thank the Center for Information Services and High Performance Computing (ZIH) at TU
Dresden for generous allocations of computer time.


\bibliographystyle{prsty}
\bibliography{t1t2t3cubic}

\begin{thebibliography}{10}
\bibitem{footnotej1j2j3}
Note that the third-neighbor coupling, $J_3$, on the square lattice
  (frustrating both $J_2$ and $J_1$) is of different nature as compared to the
  cubic lattice (where it does not frustrate $J_1$). Therefore we compare our
  cubic-lattice results to the $J_1$--$J_2$ (as opposed to $J_1$--$J_2$--$J_3$)
  square-lattice model.
\bibitem{footnoteFiniteSize}
We consider it unlikely that the NMI region represents a crossover (\ie
  critical) regime induced by finite-size effects, because (i) its width
  appears independent of the VCA cluster size and (ii) a direct transition
  between the $(\pi,\pi,\pi)$ and $(0,\pi,\pi)$ phases is expected to be of
  first order (\ie non-critical).
\bibitem{vbs}
Here we tested a columnar VBS with ordering vector $(\pi,0,0)$. Alternative
  candidate states are a plaquette VBS with $(\pi,\pi,0)$ or a cuboid VBS with
  $(\pi,\pi,\pi)$.
\end{thebibliography}

\end{document}